\documentclass[aps,twocolumn]{revtex4}
\usepackage{amsmath}
\usepackage{graphicx}
\usepackage{subeqnarray}
\begin{document}

\title{Finite Size Effect in Persistence in Random Walk}
\author{
D.Chakraborty \footnote {e-mail:tpdc2@mahendra.iacs.res.in } and
J.K.Bhattacharjee\footnote{e-mail:tpjkb@mahendra.iacs.res.in}}
\affiliation{Department of Theoretical Physics,\\
Indian Association for the Cultivation of Science,\\
Jadavpur, Calcutta 700 032, India.}

\begin{abstract}
We have investigated the random walk problem in a finite system and studied the crossover induced in 
the persistence probability by the system size.Analytical and 
numerical work show that the scaling function is an exponentially decaying
function. We consider two cases of trapping, one by a box of size  $L$ 
and the other by a harmonic trap.
Our analytic calculations are supported by numerical works.
We also present numerical 
results on the harmonically trapped randomly accelerated particle and the randomly accelerated particle with viscous drag.
\end{abstract}
\maketitle

The phenomenon of persistence has attracted a lot of interest in the recent 
years, both theoretically \cite{1}-\cite{9} as well as experimentally\cite{10}-\cite{12}.
The word "persistence" itself
conveys the meaning of survival. Associated with this survival is the survival 
probability $p(t)$. It is simply the probability that the local field 
has not yet changed its sign upto time $t$. 
For a wide range of models the survival probability 
decays as a power law, that is $\displaystyle{p(t) \sim t^{-\theta}}$, 
where $\theta$ is a new non-trivial exponent called the persistence exponent. 
Established results, exist for many 
models- Random Walk problem, Diffusion problem \cite{1},Ising Model with 
Glauber dynamics \cite{2}, Surface growth \cite{6}, Phase ordering kinetics \cite{5}.\\

In an experimental setup,however,finite size effects appear because
of the size of the apparatus and boundary effects come into play in the dynamics of the system. As a result the survival probability also depends on 
the finite size parameter. A particularly clear example of the crossover effects induced by finite size is the recent experiment on a single polystyrene sphere in a harmonic potential \cite{19}
It is this demonstration of crossover effect that has motivated us in our present work in investigating the finite size effect on the 
survival probability and how it scales with the finite size parameter.

There has been an investigation of this for the Ising system in higher dimension \cite{20} which lead to the conclusion that finite size scaling in 
the usual sense holds for persistence as well. The system that we
investigate here do not show finite size scaling in exactly that sense, although they exhibit pronounced finite size effects.
We have considered an analytically solvable model in the present work-
the case of a brownian particle confined in a box and the case of a 
brownian particle trapped by a harmonic potential. We find that the survival probability $P(t,L)$ does not have the usual scaling form $P(t,l) \sim 
t^{- \theta} f(\frac{t}{L^z})$, with $f(x) \rightarrow  constant$ for
$x \rightarrow 0$ and $f(x) \sim x^{\theta}$ for $x>>1$. Instead, we find that $P(t,L)$ can be expressed as $P(t,L)=t^{-\theta}f(\frac{t}{L^z})$
but $f(x) \rightarrow 0$ exponentially as $x \rightarrow \infty$.
This is consistent with the generic form anticipated by Redner\cite{21}.
The exponent $z$ is found to be $2.0$ as in the case of Manoj and Ray \cite{20}.

The simplest of all the models for which there exists 
an established result is the Random walk problem. 
A random walker obeys a differential equation of the form
$\displaystyle{\frac{d x(t)}{d t} = \eta (t)}$
where $\displaystyle{\eta (t)}$ is a white noise. To find the survival 
probability we ask the question whether the quantity 
$sgn[x(t)-<x(t)>]$ has changed its sign upto time $t$. 
The survival probability $P_0(T)$ in terms of the variable $\sigma = sgn(X(T))$ can be found from $A(T) = <\sigma(0) \sigma(T)>$ \cite{1}. 
In this case 
$P_0(T) = (2/\pi) sin^{-1} [exp(-\lambda T)]$.
For a random walker $\lambda = 1/2$ and the survival probability goes as
$P_0(T) \sim exp(-T/2)$ \cite{18}.
Analytical and numerical result show that the probability goes as
$p(t) \sim t^{-\frac{1}{2}}$ and the persistence
exponent in this case is
$\theta = \frac{1}{2}$ \cite{18}.

We have investigated the finite size effect 
in the random walk problem in two ways. Firstly, the random walker 
is constrained to move in a box with reflective boundaries at $x=\pm L$
The probability distribution $P(x,t)$ in this case obeys a diffusion 
equation \cite{14} with an appropriate boundary condition. 
Solution to the diffusion equation with the proper boundary 
condition gives $P(x,t)$. In the second problem, the random walker is 
trapped in a harmonic potential. Both the problems are analogous 
to each other with the identification $\omega \sim \frac{1}{L}$.
In both cases we calculate the correlator
$a(t_1,t_2)=<x(t_2)x(t_1)>$, where $x(t)$ is the value of $x$ at time $t$.
To make it a Gaussian Stationary Process (G.S.P) 
we transform $x(t)$ to $\displaystyle{\bar{X}=\frac{x(t)}{\sqrt{<x^2(t)>}}}$ 
and a suitable transformation for the time variable from $t$ to $T$.
Thus the correlator $a(t_2,t_1) \rightarrow f(|T_2-T_1|)$.
From the correlator $f(T)$ we get the survival probability $p(t)$.\\

We first consider a particle in one-dimension performing random walk. 
The equation governing the dynamics 
of the particle is given by
\begin{equation}
\label{1}
\frac{d x(t)}{d t} = \eta(t)
\end{equation}
where $x(t)$ is the displacement of the particle and $\eta(t)$ is a 
random function whose moments are given by
\begin{subeqnarray}
\label{2}
\slabel{2a}
<\eta (t)>  &=& 0 \\
\slabel{2b}
<\eta (t) \eta (t')> &=& D \delta(t-t')
\end{subeqnarray}
where $D$ is the diffusion coefficient.
In the present problem we confine the motion of the particle within a 
cage with boundaries at $x=\pm L$. The boundary of the cage is 
reflective that is upon reaching the boundary
the particle is reflected to the nearest lattice site.\\

The probability $P(x,t)$ that the coordinate is $x$ at a time $t$ starting from $x=0$ at $t=0$ obeys the diffusion equation
\begin{equation}
\label{3}
\frac{\partial P(x,t)}{\partial t} = D \frac{\partial^2 P(x,t)}{\partial^2 x}
\end{equation}
with proper boundary condition. 
Since the particle is reflected from the boundary,
the particle current $x=\pm L$ must be zero. Thus at the boundary 
\begin{equation}
\label{4}
-D \frac{\partial P(x,t)}{\partial x} \biggr |_{x=\pm L} = 0
\end{equation}
The solution to Eq.(\ref{3}) with the boundary condition Eq.(\ref{4})
can be written as
\begin{equation}
\label{5}
P(x,t)=\cos(\frac{n \pi x}{L})\phantom{1} e^{-\frac{n^2 \pi^2 Dt}{L^2}}
\end{equation}
The complete form of the probability $P(x,t)$ taking into consideration 
the normalization can be written down as
\begin{equation}
\label{6}
P(x,t)=\frac{1}{2L}+\frac{1}{L} \sum_{n=1}^{\infty} 
\cos \biggr (\frac{n \pi x}{L}\biggr )  \phantom{1} 
e^{-\frac{n^2 \pi^2 D t}{L^2}}
\end{equation}
The average of the square of the displacement is given by
\begin{eqnarray}
\label{7}
\nonumber
<(\triangle x)^2> &=& \frac{L^2}{3} + \frac{4L^2}{\pi^2} 
\sum_{n=1}^{\infty} \frac{1}{n^2}
\cos(n \pi) e^{-\frac{n^2 \pi^2 D t}{L^2}}\\
\end{eqnarray}
This is the exact answer. As expected it exhibits finite size scaling
and can be cast in the form 
\begin{equation}
\label{8}
<(\triangle x)^2> = \frac{L^2}{3} g(\frac{t}{L^2})
\end{equation}
such that $g(\frac{t}{L^2}) \propto \frac{t}{L^2}$ for $\frac{t}{L^2} \rightarrow 0$ (i.e infinite system size) and 
$g(\frac{t}{L^2}) \rightarrow 1$ for $\frac{t}{L^2} \rightarrow \infty$
that is the extreme case of finite system.

To see this the sum in Eq.(\ref{7}) can be decomposed as
\begin{eqnarray}
\label{9}
\nonumber
<(\triangle x)^2>=\frac{L^2}{3}-\frac{4L^2}{\pi^2} 
\biggr (\sum_{\textrm{n odd}} \frac{1}{n^2}
\phantom{1} e^{-\frac{n^2 \pi^2 D t}{L^2}} \\
+\sum_{\textrm{n even}} \frac{1}{n^2}
\phantom{1} e^{-\frac{n^2 \pi^2 D t}{L^2}} \biggr )
\end{eqnarray}
In the limit $L \rightarrow \infty$, all the modes in the summation
of Eq.(\ref{7}) must be considered. It is then easy to see that
\begin{equation}
\label{10}
<(\triangle x)^2> \rightarrow  \frac{L^2}{3} \phantom{3} 
\textrm{for $t \rightarrow \infty$ and $L$ finite}\\
\end{equation}
and taking the opposite limit we find that
\begin{equation}
\label{11}
<(\triangle x)^2>  \rightarrow  2Dt \phantom{3}
\textrm{for $L \rightarrow \infty$ and $t$ finite}\\
\end{equation}
as expected. Keeping in mind our future need where $f(x) \rightarrow 1$
as $x \rightarrow 0$ but decays very fast for $x>>L$, we will express
Eq.(\ref{7}) as an approximate which is easy to handle.
This is done
using the Euler-Maclaurin sum formula for the two sums in Eq.(\ref{9}).

Keeping only $\displaystyle{e^{-\frac{\pi^2 D t}{L^2}}}$ from among 
the different exponential decays and working to $O(\frac{t}{L^2})$ 
the coefficient of $\displaystyle{e^{-\frac{\pi^2 D t}{L^2}}}$
the crossover function can be written as 
\begin{eqnarray}
\label{12}
\nonumber
<(\triangle x)^2>=\frac{L^2}{3}-\frac{2L^2}{\pi^2} 
\int_{1}^{2} dn \frac{1}{n^2} \phantom{1} 
e^{-\frac{n^2 \pi^2 D t}{L^2}} \\
+ \biggr ( \frac{1}{\pi^2}-\frac{1}{3} \biggr)\biggr 
(1+\frac{D \pi^2 t}{L^2}
\biggr )e^{-\frac{\pi^2 D t}{L^2}}
\end{eqnarray}
A more drastic approximation yields the expression 
\begin{equation}
\label{13}
<(\triangle x)^2>=\frac{\frac{d^2 x(t)}{d t^2}L^2}{3}\biggr[1-\biggr(1+(\pi^2 -6)\frac{Dt}{L^2}
 \biggr)\phantom{1}
e^{-\frac{\pi^2 D t}{L^2}}\biggr]
\end{equation}
The part in the bracket is the approximation for the function 
$g(\frac{t}{L^2})$.
For $\frac{t}{L^2} \rightarrow \infty $, Eq.(\ref{13}) correctly reduces 
to $\frac{L^2}{3}$, while for $\frac{t}{L^2} << 1$, we expand Eq.(\ref{13})
to obtain
\begin{equation}
\label{14}
<(\triangle x)^2>=2Dt \biggr [1-\gamma \frac{Dt}{L^2} \biggr ]
\end{equation}
where 
\begin{equation}
\label{15}
\gamma=\pi^2(1-\frac{\pi^2}{12})
\end{equation}
The leading term in Eq.(\ref{14}) is the correct limit for the unbounded
system and the second term is the first correction for finite L.

We now proceed to calculate the correlator $<x(t_2)x(t_1)>$ for the dynamics
of Eq.(\ref{1}) keeping in mind the approximations used in arriving at Eq.(\ref{13}) for $<(\triangle x(t))^2>$. Considering the equation for the 
probability distribution (Eq.(\ref{3}), we write down the expression for 
$P(x_2,t_2;x_1,t_1)$ the probability of finding a value $x_2$ at
$t=t_2$ if the value was $x_1 $ at $ t=t_1$. We note the exact result 
\begin{eqnarray}
\label{16}
\nonumber
\int [x(t_2)-x(t_1)]^2 P(x_2,t_2;x_1,t_1) dx_2dx_1 \\
=<(\triangle x(t_2-t_1))^2>
\end{eqnarray}
whence
\begin{eqnarray}
\label{17}
\nonumber
<x^2(t_2)>+<x^2(t_1)>-2<x(t_2)x(t_1)>\\
=<(\triangle x(t_2-t_1))^2>
\end{eqnarray}
and we can now use Eq.(\ref{13}) to calculate $a(t_1,t_2)=<x(t_2)x(t_1)>$.
To obtain a Gaussian Stationary Process, it is necessary to calculate the 
correlation of $\bar{X}=\frac{x(t)}{\sqrt{x^2(t)}}$ and this leads to 
complicated looking expression. To express the final answer in a particularly
simple form, we use the regime $\frac{t}{L^2}<<1$ and then exponentiate to find
\begin{eqnarray}
\label{18}
<\bar{X}(t_2)\bar{X}(t_1)>=\sqrt{\frac{t_1}{t_2}}\phantom{1}
e^{-\gamma \frac{D(t_2-t_1)}{2L^2}}
\end{eqnarray}
This exponentiation is predicted by the generic form anticipated by Redner. In the process of our calculation, we find the numerical prefactors which are not present in the general arguments of Redner. For comparison with the numerical simulations, these prefactors are essential.
We now perform the transformation in time 
$t \rightarrow T=lnt + \frac{\gamma Dt}{L^2}$. 
The correlator $f(T_2,T_1)$in the transformed variables is
\begin{equation}
\label{19}
<\bar{X}(T_2)\bar{X}(T_1)>=e^{-1/2(T_2-T_1)}
\end{equation}
The process is now a Gaussian Stationary Process(G.S.P).
The survival probability is now given by
\begin{eqnarray}
\label{20}
p(t)= t^{-1/2}\phantom{1}e^{-\frac{\gamma Dt}{2L^2}}
=t^{-1/2} f(\frac{t}{L^2})
\end{eqnarray}

To test Eq.(\ref{20}), we have calculated $p(t)$ numerically.
\begin{equation}
\label{21}
t^{1/2}p(t)=e^{-\frac{\gamma Dt}{2L^2}}
\end{equation}
We expect the semi-log plot of $t^{-1/2}p(t)$ vs $\displaystyle{\frac{Dt}{L^2}}$ to be straight line with
a slope of $\bar{\gamma}=\gamma/2$.
The value of $\bar{\gamma}$ obtained from our calculations is
\begin{equation}
\label{22}
\bar{\gamma}=\frac{\pi^2}{2}(1-\frac{\pi^2}{12})=0.8761
\end{equation}
Numerical simulation of the process was done using various values of L. 
The probability was obtained by averaging over $10^6 $ configurations.
Numerically obtained value of $\bar{\gamma}$ is 0.9482. This discrepancy can
be attributed to the approximate form of Eq.(\ref{18}).
A semi-log plot of
$t^{\frac{1}{2}}p(t)$ vs $\frac{Dt}{L^2}$ is shown in Fig 1.
This clearly shows the validity of Redner's generic form and the reasonableness of our approximations in arriving at the numerical value for $\bar{\gamma}$ and the fact that $z=2$.

\begin{figure}
\centering
\includegraphics[width=8cm,height=6cm]{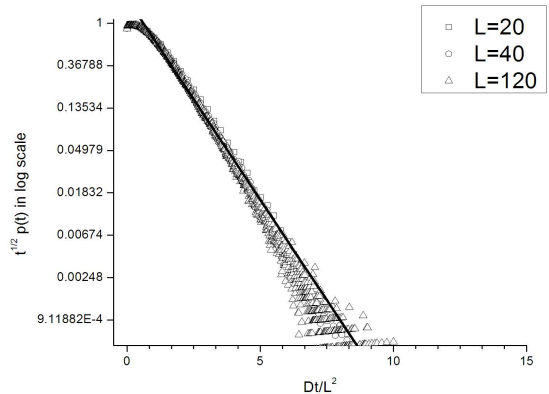}
\caption{Semi-log plot of $t^{\frac{1}{2}}p(t)$ vs $\frac{Dt}{L^2}$. 
The straight line is the best fit for the linear part of the curve.}
\end{figure}

We next consider a particle trapped in a harmonic potential 
and acted upon by a random force. Instead of sharp boundaries, we now have the potential confining the particle. If the confining length is $L$, then on dimensional grounds we expect $\omega \sim 1/L$
This is the set up of Ref \cite{19} except there the inertial effect cannot be ignored.
The equation of motion is 
\begin{equation}
\label{23}
\frac{d x(t)}{dt}+\omega^2 x=\eta (t)
\end{equation}
where $\eta (t)$ is the random noise whose moments are given by
Eq.(\ref{2a}) and Eq.(\ref{2b}).
The expression for $x(t)$ then becomes
\begin{equation}
\label{24}
x(t)=e^{-\omega^2 t}\int_{0}^{t}e^{\omega^2 t'} \eta (t') dt'
\end{equation}
The correlator $<x(t_1)x(t_2)>$ is given by
\begin{eqnarray}
\label{25}
\nonumber
<x(t_1)x(t_2)>&=&e^{-\omega^2 (t_1+t_2)}
\int_{0}^{t_1}\int_{0}^{t_2}
e^{\omega^2 (t'_1+t'_2)}\\
&<&\eta (t'_1)\eta (t'_2)> dt'_1 dt'_2
\end{eqnarray}
Using Eq.(\ref{2b}) we have
\begin{equation}
\label{26}
<x(t_1)x(t_2)>=\frac{\epsilon}{2\omega^2}[e^{-\omega^2 (t_1-t_2)}-
e^{-\omega^2 (t_1+t_2)}]
\end{equation}
The correlator in the new scaled variable 
$x(t) \rightarrow \bar{X}(t)=\frac{x(t)}{\sqrt{<x^2(t)>}}$ 
has the form
\begin{equation}
\label{27}
<\bar{X}(t_1)\bar{X}(t_2)>=e^{-\frac{\omega^2}{2}(t_1-t_2)}\biggr[
\frac{sinh(\omega^2 t_2)}{sinh(\omega^2 t_1)} \biggr]^{\frac{1}{2}}
\end{equation}
Writing $\displaystyle{e^{T}=\frac{1}{\omega^2}e^{\omega^2 t} sinh(\omega^2 t)}$
we have
\begin{equation}
\label{28}
f(T_1,T_2)=<\bar{X}(T_1)\bar{X}(T_2)>=e^{-\lambda (T_1-T_2)}
\end{equation}
where $\lambda=\frac{1}{2}$. The process is now a gaussian stationary process. 
The survival probability can now be written as
\begin{equation}
\label{29}
p(T)=e^{-\lambda T}
\end{equation}
and in real time the survival probability is
\begin{eqnarray}
\label{30}
\nonumber
p(t)&=&[\frac{1}{\omega^2}\phantom{1}
e^{\omega^2 t}\phantom{1}sinh(\omega^2 t)]^{-\frac{1}{2}}\\
&=&\frac{\omega e^{-\omega^2 t/2}}{\sqrt{sinh(\omega^2 t)}}
=\frac{1}{t^{1/2}} \phantom{2} f(\omega^2 t)
\end{eqnarray}
where
\begin{equation}
\label{31}
f(x)=\sqrt{\frac{x}{sinh(x)}} \phantom{2} e^{-x}
\end{equation}
As stated before $z=2$ and $f(x) \rightarrow 1$ as $ x \rightarrow 0$,
while $f(x) \rightarrow 0 $ as $ x \rightarrow \infty$.
For $\omega \rightarrow 0$ the above expression for probability reduces to the normal random walk
problem and the probability $p(t)$ goes as $t^{-\frac{1}{2}}$.
The numerical data is obtained for three values of $\omega$. For $t<1/\omega^2$ the estimated value of the exponent by fitting the log-log plot with a straight line is found to be $\theta=0.5055$.
\begin{figure}
\centering
\includegraphics[width=8cm,height=6cm]{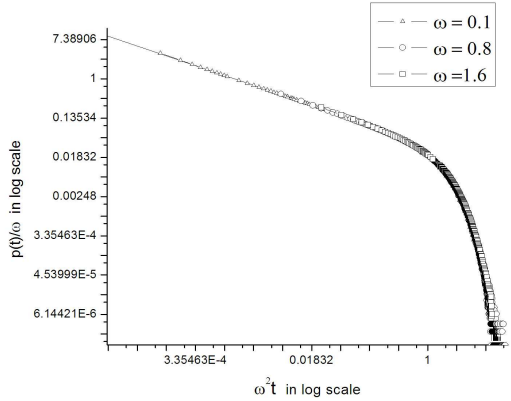}
\caption{Log-log plot of $p(t)$ vs $\omega^{2}t$. The line shows the
function $\frac{e^{-\frac{\omega^2 t}{2}}}{\sqrt{sinh(\omega^2 t)}}$}
\end{figure}

Finally we present numerical studies of the survival probability for a
harmonically trapped randomly accelerated particle and for randomly accelerated
particle with viscous drag. The persistence exponent for the randomly accelerated
particle is $\theta=0.25$ \cite{15}-\cite{17}.
The equation of motion for the
particle is
\begin{eqnarray}
\label{32}
\frac{d^2 x}{d t^2}+\omega^2 x= \eta(t)
\end{eqnarray}
where $\eta(t)$ is a gaussian white noise with correlator given by Eq.(\ref{2a})
and Eq.(\ref{2b}). Rescaling $\tau=\omega t$ we see that for $\tau<<1$
or $t<<1/\omega$ the first term domainates and the equation is motion is that
of a randomly accelerated particle. A plot of survival probability vs time is
shown in Fig. 3. With $\omega$ behaving as $\frac{1}{L}$ as noted above, this corresponds to $z=1$ for the dynamic component.
The survival probability is obtained by averaging over
$10^5$ configurations.
\begin{figure}
\centering
\includegraphics[width=8cm,height=6cm]{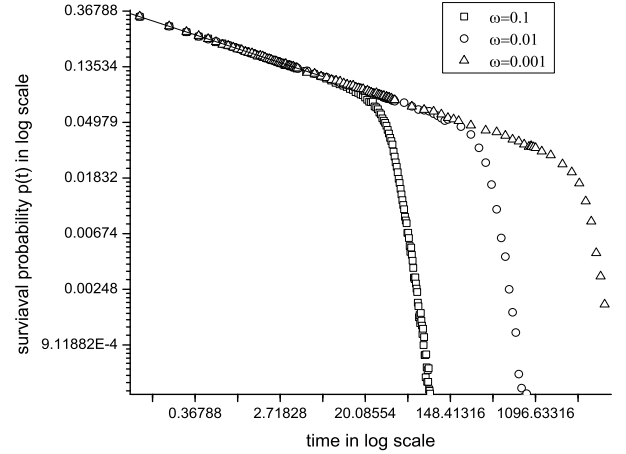}
\caption{log-log plot of survival probability vs time. The solid line
is the best fit line. The estimated persistence exponent by best fit is
$\theta=0.2502$.}
\end{figure}

For the randomly accelerated particle with viscous drag the equation of motion is
\begin{eqnarray}
\label{33}
\frac{d^2 x}{d t^2}+\Gamma \frac{d x}{d t}= \eta(t)
\end{eqnarray}
This is the crossover which is of particular interest in the experiment of
Ref \cite{19}. As can be seen from our results the cross over in the form of the correlation function occurs at $t \sim \frac{1}{\Gamma}$. 
In this case however two regimes exists. For $t<<1/\Gamma$ the equation
of motion is that of a randomly accelerated particle with the first term
dominating and for $t>>1/\Gamma$ the second term domainates and the equation
of motion is that of a random walker. Thus the survival probability also
shows a crossover from the randomly accelerated regime to random walk regime.
The estimated values of the exponents in the two regimes is tabulated below.\\
\begin{tabular}{|c|c|c|}
\hline
value of $\Gamma$ & $\theta$ for $t<1/\Gamma$ & $\theta$ for $t>1/\Gamma$\\
\hline
0.8 & 0.2638 & 0.4970 \\
\hline
0.1 & 0.2510 & 0.4888 \\
\hline
0.01 & 0.2538 & 0.4797 \\
\hline
\end{tabular}

Numerically obtained values of survival probability is plotted against time
in Fig. 4. The survival probabilities is obtained by averaging over $10^5$ 
configurations.
One of the important finding of Ref \cite{19} was that in the realistic situation of the experiment the crossover in $<(\triangle x)^2>$ occurred for $t>>\frac{1}{\Gamma}$. This was attributed to a memory dependent damping term. In a future work, we will explore the effect of it on the persistence problem.
\begin{figure*}
\centering
\includegraphics[width=8cm,height=6cm]{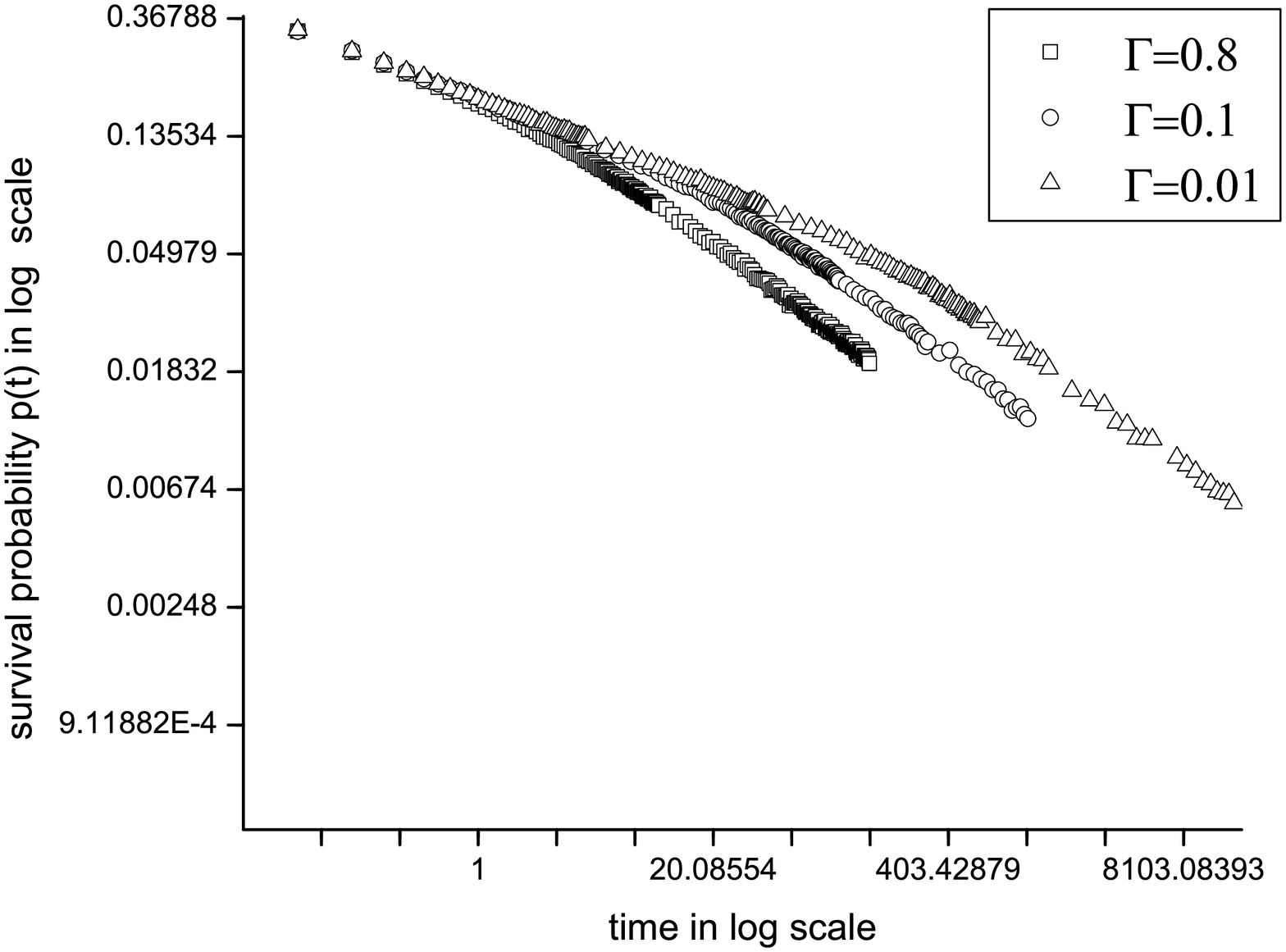}
\caption{log-log plot of survival probability vs time.}
\end{figure*}

{\bf{\large{Acknowledgement:}}}\\
D.C acknowledges Council for Scientific and Industrial Research, 
Govt. of India for financial support (Grant No.- 9/80(479)/2005-EMR-I).

\end{document}